\documentclass[showpacs,aps,twocolumn]{revtex4}
\usepackage{graphicx}
\usepackage{bm}

\begin{document}
\title{General Relativistic effects on the conversion of nuclear 
to two-flavour quark matter in compact stars}
\author{Abhijit Bhattacharyya}
\email{abphy@caluniv.ac.in}
\affiliation{Department of Physics, University of Calcutta, 
92, A. P. C. Road, Kolkata - 700009; INDIA}
\author{Sanjay K. Ghosh}
\email{sanjay@bosemain.boseinst.ac.in}
\affiliation{Centre for Astroparticle Physics \&
Space Science and Department of Physics, Bose Institute, 
93/1, A.P.C Road, Kolkata - 700009, INDIA}
\author{Ritam Mallick}
\email{ritam@bosemain.boseinst.ac.in}
\affiliation{Centre for Astroparticle Physics \&
Space Science and Department of Physics, Bose Institute, 
93/1, A.P.C Road, Kolkata - 700009, INDIA}
\author{Sibaji Raha}
\email{sibaji@bosemain.boseinst.ac.in}
\affiliation{Centre for Astroparticle Physics \&
Space Science and Department of Physics, Bose Institute, 
93/1, A.P.C Road, Kolkata - 700009, INDIA}
\begin{abstract}
We investigate the General Relativistic (GR) effects on the conversion 
from nuclear to two-flavour quark matter in compact stars, both static 
as well as rotating. We find that GR effects lead to qualitative 
differences in rotating stars, indicating the inadequacy of 
non-relativistic (NR) or even Special Relativistic (SR) treatments for these cases. 
\end{abstract}
\maketitle

Strange Quark Matter (SQM), consisting of approximately equal numbers of 
up ({\it u}), down ({\it d}) and strange ({\it s}) quarks, is the putative 
{\it true} ground state of strong interaction \cite{key-1}, a conjecture 
supported by model calculations for certain ranges of values for strange 
quark mass and strong coupling constant \cite{key-2}. There have been 
concerted efforts at confirming the existence of Quark-Gluon Plasma (QGP) 
and SQM, though transiently, in ultra relativistic collisions. On the other 
hand, QGP and SQM could naturally occur in the cores of compact stars, 
where central densities of about an order of magnitude higher than the 
nuclear matter saturation density are expected. Given the very low 
strangeness fraction in normal nuclear matter (NM), it is almost inevitable 
that a transition from nuclear (hadronic) to quark matter should proceed 
through a conversion to an initial stage of (metastable) two flavour quark 
matter, which should decay to the stable SQM. Thus, neutron stars with 
sufficiently high central densities ought to get converted to strange, 
or at least hybrid, stars. These transitions could have observable 
signatures in the form of a jump in the breaking index and gamma ray 
bursts  \cite{key-3,key-4}. On the other hand a full quark star may explain 
the phenomena of observed quasi periodic oscillations \cite{key-4a}.
\par
There are several plausible scenarios where neutron stars could convert
to quark stars, through a "seed" of external SQM \cite{key-5}, or 
triggered by the rise in the central density due to a sudden spin-down 
in older neutron stars \cite{key-6}. Several authors have studied the
conversion of nuclear matter to strange matter under different 
assumptions \cite{key-7,key-8,key-9,key-10,key-11,key-12,key-13,
key-14,key-15,key-16,key-17}. They have been summarized in a recent
work of ours \cite{key-18} and for the constraint of space, we do not
repeat them here, except to mention that Tokareva {\it et al} \cite{key-14} 
have lately modelled the hadron to SQM conversion as a single 
step process, arguing that the mode of conversion would vary 
with the temperature as well as the value of bag constant. Berezhiani
{\it et al} \cite{key-15}, Bombaci {\it et al} \cite{key-16} and Drago 
{\it et al} \cite{key-17} suggested that the formation of SQM may be 
delayed, if the deconfinement process takes place through a
first order transition \cite{key-19} so that the purely hadronic
star can spend some time as a metastable object.
\par
In our recent work \cite{key-18}, we have argued that the conversion process 
is a two step process. The first process involves the deconfinement of 
nuclear to two-flavour quark matter and the conversion process takes some 
milliseconds to occur. The second process deals with the conversion of excess 
down quarks to strange quarks, which occurs via weak interaction, forming a 
stable SQM and the time taken for this process to occur is of the order of 
100 seconds. This work was restricted to the case of a static neutron star and 
only a SR treatment. To the best of our knowledge, GR 
effects in such processes have not yet been addressed in the literature.
In this letter, we consider GR effects in both static and rotating stars 
and find qualitatively new results.
\par
As in \cite{key-18}, we use the Nonlinear Walecka model for the nuclear matter 
euqation of state (EOS). For the sake of brevity, we do not repeat the details 
here. Suffice it to say, the star is assumed to be composed of only nucleons. 
The metric describing the structure of the star, is given by \cite{key-20}
\begin{eqnarray}
ds^2 = -e^{\gamma+\rho}dt^2 + e^{2\alpha}(dr^2+r^2d\theta^2) + \nonumber \\ 
e^{\gamma-\rho}r^2 sin^2\theta(d\phi-\omega dt)^2
\end{eqnarray}
The four gravitational potentials $\alpha, \gamma, \rho$ and $\omega$ are 
functions of $\theta$ and $r$ only. Once these potentials are known, one can 
calculate the observed properties of the star. We solve the Einstein's 
equations for the three potentials $\gamma, \rho$ and $\omega$, using
Green's function technique \cite{key-21,key-3} and determine the fourth 
potential $\alpha$ from the other three potentials. The solution of the star 
is obtained from the {\bf 'rns'} code \cite{key-23}, which requires the EOS 
and a central density as inputs and returns various gravitational potentials 
and hydrodynamic parameters as outputs. We solve this code for a static as 
well as for rotating stars with different velocities, upto the mass-shedding 
limit or the Keplerian velocity. The results are, of course, as expected; 
rotation induces a change in the star to an oblate-spheroid shape.
\par
To study the GR effects on the conversion of nuclear to two flavour quark 
matter, we heuristically assume the existence of a combustive phase transition 
front of infinitesimal thickness, and study the outward propagation of the 
front, using GR hydrodynamical equations of motion. First we consider a 
non-rotating neutron star, where the geometry of the problem 
is one-dimensional and then extend our calculation to a rotating neutron star, 
where an extra dimension is needed. Let us assume that the conversion front 
is generated at the center of the star, and it propagates outwards through 
the star with a certain velocity, dictated by the initial conditions and 
hydrodynamical equations, converting the nuclear matter to two-flavour quark 
matter. Employing the conservation conditions \cite{key-24,key-25} and 
further employing the entropy condition \cite{key-26}, we determine the flow 
velocity of matter in the two phases, given by
\begin{equation}
v_{1}^{2}=\frac{(p_{2}-p_{1})(\epsilon_{2}+p_{1})}{(\epsilon_{2}
-\epsilon_{1})(\epsilon_{1}+p_{2})},
\end{equation}
and 
\begin{equation}
v_{2}^{2}=\frac{(p_{2}-p_{1})(\epsilon_{1}+p_{2})}{(\epsilon_{2}
-\epsilon_{1})(\epsilon_{2}+p_{1})}.
\end{equation}
It is possible to classify the various conversion mechanisms by comparing
the velocities of the respective phases with the corresponding velocities
of sound, denoted by $c_{si}$, in these phases. For the conversion to be 
physically possible, velocities should satisfy an additional condition, 
namely, $0\leq v_{i}^{2}\leq 1$; here we have used natural units $\hbar = c = K = 1$. 
If we plot different velocities against 
baryon number, we get curves from which we could calculate the initial 
velocity of the front at the centre of the star \cite{key-18}. Starting 
with this initial velocity, we investigate the evolution of the front with 
time under GR effect; to this end, we first have to derive the appropriate 
continuity and Euler's equations for the metric [eq. (1)].  Treating both 
nuclear and strange matters as ideal fluids, the system is governed, 
together with the metric and the EOS, by the Einstein's equation 
${R_i}^k - \frac{1}{2}{\delta_i}^k R = \kappa {T_i}^k$ and the equation of motion 
${T^k}_{i;k}=0$  \cite{key-27}.
\par
For adiabatic motion, we have,
\begin{eqnarray}
\frac{d(\varpi u_i)}{ds} + \frac{\partial \varpi}{\partial x^i} = 
\frac{\varpi}{2} u^k u^l \frac{\partial g^{kl}}{\partial x^i} + 
T \frac{\partial \sigma}{\partial x^i} \\ 
\frac{\partial}{\partial x^k} ( \frac{u^k}{V}\sqrt{-g})=0 ; \hspace {0.3 in}  \frac{\partial \sigma}{\partial s}=0
\end{eqnarray}
where $u^i$ signifies the four velocity, $\varpi$ and $\sigma$ are the enthalpy 
and entropy, respectively. For an isentropic process, {\it i.e.} $\sigma=constant$, 
the term $T \frac{\partial \sigma}{\partial x^i}$ goes to zero.
\par
The above two equations are the starting point for deriving the appropriate
continuity and Euler's equations. If we want the equation for 
$\frac{\partial v}{\partial r}$, we only need to consider the variation of the 
gravitational potentials with $r$ and the terms containing partial derivatives 
w.r.t $\theta$ can be neglected. For our metric,  $\sqrt{-g}$ comes out to be
$\sqrt{-g}= e^{\gamma + 2\alpha}r^2 sin\theta $; the four vectors are also 
calculated from the metric. To derive the hydrodynamical equations, we use 
'auxiliary' time $d\tau$ instead of ordinary time $dt$, defined as
$ d\tau=\frac{\sqrt{g_{rr}}}{\sqrt{g_{tt}}} dt $.
\par
Calculating these values from the metric and putting them in eqn. (4) and (5), 
we get the continuity and the Euler's equations, given by
\begin{eqnarray}
\frac{1}{\varpi}(\frac{\partial\epsilon}{\partial\tau}
+v\frac{\partial\epsilon}{\partial r})+
\frac{1}{W^{2}}(\frac{\partial v}{\partial r}
+v\frac{\partial v}{\partial\tau})+\frac{2v}{r}+\frac{v}{r}cot\theta  \nonumber\\
= -v (\frac{\partial\gamma}
{\partial r} + \frac{\partial \alpha}{\partial r})\\
\frac{1}{\varpi}(\frac{\partial p}{\partial r}+
v\frac{\partial p}{\partial\tau})+
\frac{1}{W^{2}}(\frac{\partial v}{\partial\tau}+
v\frac{\partial v}{\partial r}) \nonumber \\
=\frac{1}{2}(A\frac{\partial\gamma}
{\partial r} + B\frac{\partial \rho}{\partial r}+C\frac{\partial \omega}{\partial r}+E),
\end{eqnarray}
where, $v$ is the r.m.s velocity and $W$ is the inverse of the Lorentz factor. 
This system of equations differs from its SR counterpart in the 
appearance of the gravitational potentials, describing the gravitational forces 
acting on the fluid in its own gravity field. Once the two basic equation are 
set, we now proceed exactly as in \cite{key-18}. We define $v$ as the front 
velocity in the nuclear matter rest frame and $n=\frac{\partial p}
{\partial\epsilon}$ is the square of the effective sound speed in the medium. 
$d\tau$ and $dr$ are connected by the relation
\begin{equation}
\frac{d r}{d \tau}=vG
\end{equation}
where $G$ is given by
\begin{equation}
G=\sqrt{\frac{e^{\gamma+\rho}-e^{\gamma-\rho}r^2\omega^2 sin^2\theta}{e^{2\alpha}}}
\end{equation}
The other parameters of the equation are defined as 
\begin{eqnarray}
A=\frac{v\omega r sin\theta}{C1}-1; \hspace {0.1 in} E=\frac{2\omega^2 r sin\theta}{C1}+\frac{2\omega^2 e^{\gamma-\rho} r sin\theta}{A1}; \nonumber\\
B=\frac{B1}{A1}-\frac{v\omega r sin\theta}{C1}; \hspace {0.1 in} C=\frac{2\omega e^{\gamma-\rho} r^2 sin\theta}{A1}+\frac{vrsin\theta}{C1} \nonumber 
\end{eqnarray}
where 
\begin{eqnarray}
A1=e^{\gamma+\rho}-e^{\gamma-\rho}r^2\omega^2 sin^2\theta; \nonumber \\ 
B1=-e^{\gamma+\rho}-e^{\gamma-\rho}r^2\omega^2 sin^2\theta;  \nonumber \\
C1=\sqrt{r^2\omega^2sin^2\theta-e^{2\rho}}; \nonumber 
\end{eqnarray}
\par
After a bit of algebra, we get a single differential equation for $v$: 
\begin{equation}
\frac{\partial v}{\partial r} = \frac{W^2v[K+K1]}{2[v^2(1+G)^2-n(1+v^2G)^2]}.
\end{equation}
where,
\begin{eqnarray}
K=2n(1+v^2G)({\frac{\partial \gamma}{\partial r} + 
\frac{\partial \alpha}{\partial r}+\frac{2}{r} +\frac{cot\theta}{r}
}) \nonumber \\
K1=(1+G)(A\frac{\partial \gamma}{\partial r} + B\frac{\partial \rho}{\partial r}
+C\frac{\partial \omega}{\partial r}+E) \nonumber
\end{eqnarray}
$\omega=0$ and $sin\theta=1$ in this equation yield the equation for the static star and if we put all potentials equal to zero, we recover our equation for 
the SR case \cite{key-18}.
\par
Integrating eq. (10) over r from the center to the surface, we obtain the 
propagation velocity of the  front along the radial direction. The initial 
condition, {\it i.e}, the velocity at the center of the star, should be zero 
from symmetry considerations. On the other hand, the $1/r$ dependence of the 
$dv/dr$ in eqn. (10) suggests a steep rise in velocity near the center of the 
star. 
\par
Our calculation proceeds as follows. Having constructed the density profile of 
the star for a fixed central density, eqns. (2) and (3) specify the respective 
flow velocities $v_{1}$ and $v_{2}$ of the nuclear and quark matter in the 
rest frame of the front, at a radius infinitesimally close to the center of 
the star. This would give us the initial velocity of the front ($-v_{1}$), 
at that radius, in the nuclear matter rest frame. With this, we integrate 
eqn. (10) outwards along the radius of the star. The solution gives the 
variation of the velocity with position as a function of the time of arrival 
of the front. Using this velocity profile, we can calculate the time required
to convert the whole star using the relation $\frac{dr}{d\tau}=vG$.
\begin{figure}[h]
\vskip 0.2in
\centering
\includegraphics[width=2.75in]{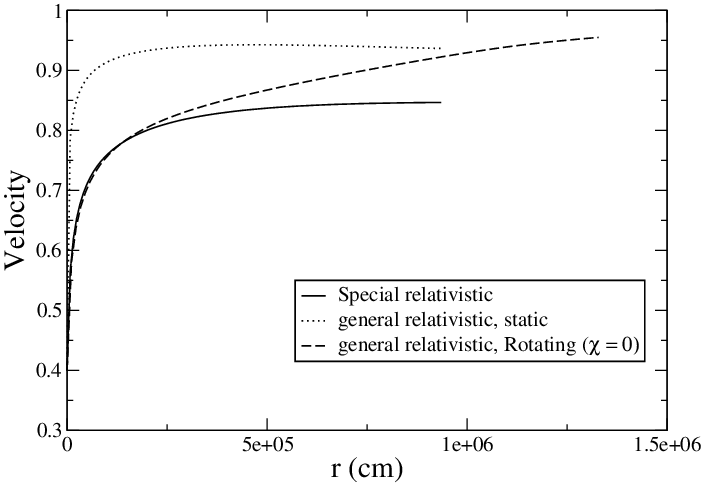}   
\caption{Variation of velocity of the front along the radial direction of
 the star for three different cases, namely SR, non-rotating GR and rotating GR.}
\end{figure}
\par
For a static star, being spherically symmetric, the problem is rather simple; 
for a rotating star, however, the asymmetry has to be taken care of. For that, 
we would have to introduce a new parameter $\chi=cos\theta$, along the 
vertical axis of the star. We start our calculation by choosing the central 
density of the star to be $7$ times the nuclear matter density, for which the 
Keplerian velocity of the star is $0.67 \times 10^{-4} sec^{-1}$ (the 
rotational velocities given in fig. (2) are all in units of $10^{-4} sec^{-1}$). 
For this central density, the initial velocity of the front comes out to 
be $0.45  $. For a complete understanding of the GR effect, comparison 
with respect to SR is desirable. After solving the differential equation for 
static and rotating star, we have plotted in fig. (1) the propagation velocity 
of the front along the radial direction of the star for three cases. The unbroken 
curve is for the 
SR case, the broken curve for non-rotating GR case and the dotted curve for 
the rotating GR case with $\chi=0$, {\it i.e} at the equator. As expected, 
for all the three cases, the velocity shoots up near the center of the star 
and saturates at a certain velocity for large radii. This is due to the 
asymptotic behaviour of the differential equations. It can also be clearly 
seen that the GR effect increases the velocity of the front considerably 
(maximum by $30 \%$) and the effect is most pronounced for the static case. 
The rotational effect of the star seems to suppress the GR effect and 
therefore the velocity of the front decreases. The result becomes clearer if 
we look at fig. (2) where we have plotted the front velocity with equatorial 
radius for different rotational velocities; as the rotational velocity 
increases, the velocity of the front decreases.

\begin{figure}[t]
\vskip 0.2in
\centering
\includegraphics[width=2.75in]{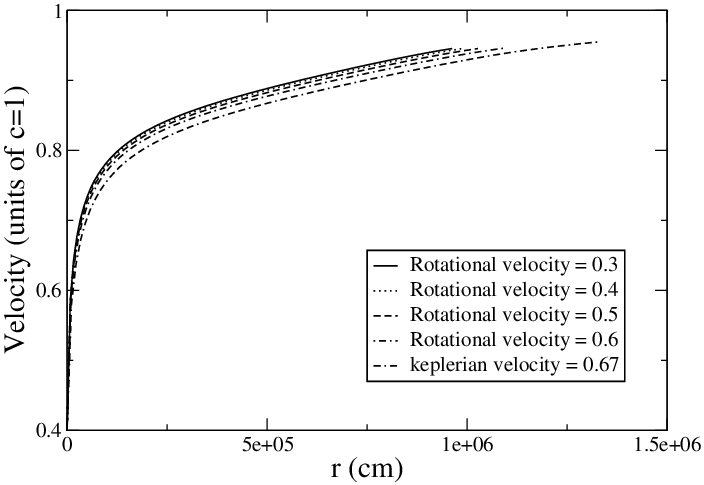}   
\caption{Variation of velocity of the front along the radial direction for
 different rotational velocity of the star.}
\end{figure}

\begin{figure}[b]
\centering
\includegraphics[width=2.75in]{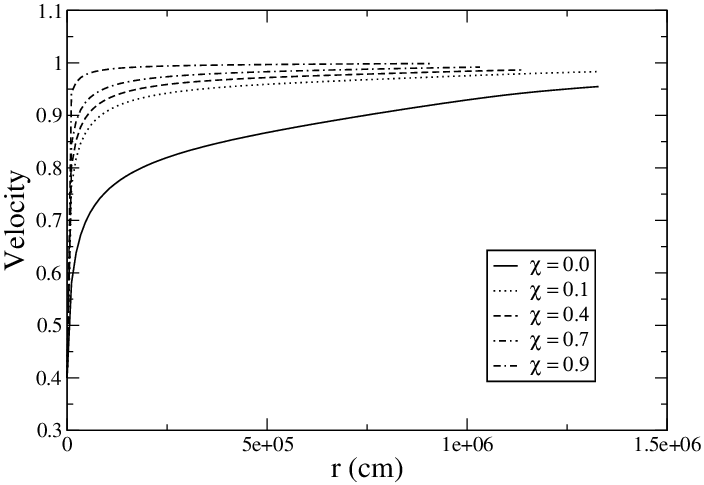}   
\caption{Variation of velocity of the front along the radial direction for
 different  $\chi$.}
\end{figure}

\begin{figure}[t]
\vskip 0.2in
\centering
\includegraphics[width=2.75in]{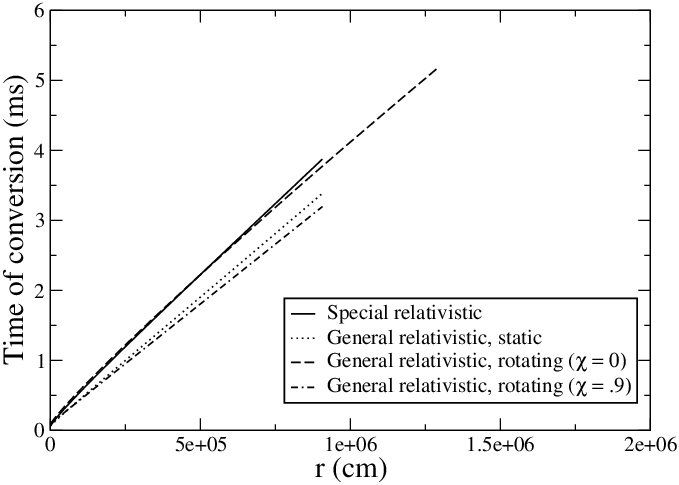}   
\caption{Variation of time of arrival of the front at certain radial distance
 for different cases.}
\end{figure}

\par
From fig. (3), we find that the front velocity is maximum along the polar 
direction and minimum along the equator. This is quite an interesting result 
as the equator has the maximum length and the front takes the maximum time 
to convert nuclear matter to two-flavour quark matter in that direction than 
towards the pole. Therefore, at any particular instant of time, we may have a 
situation where the polar part of the star has been converted while along the 
equatorial direction, the front is still propagating. The result is interesting 
but not unexpected as at the pole the EOS is much steeper than at the equator 
(and we have used the slope of the EOS, $'n'$ to calculate the front velocity). At some distance from the center of the star, the propagation front breaks up 
into several distinct fronts propagating with different velocities in different directions. 
\par
From fig. (4) we can see that the time taken by the conversion front to 
convert the neutron star to two-flavour quark star is of the order of few $ms$. The static star takes the minimum time ($3.3 ms$) whereas the rotating star 
takes the maximum time ($5.1 ms$) due to the enlarged equatorial radius. The 
polar part of the star needs much lesser time for conversion ($3.1 ms$), 
even less than static star, as its radius gets compressed.
\par
To summarize, we have shown in this letter that the conversion of nuclear 
matter to quark matter in compact stars, especially rotating stars which 
are more realistic than static stars, is strongly affected by GR effects. 
The emergence of different conversion fronts, propagating with different 
velocities along different radial directions, is a novel finding, not 
anticipated by Newtonian or SR calculations. It remains to be explored whether 
the incorporation of dissipative effects materially changes the conclusion. 
Intuition would point to a dampening of motion under dissipation, but
the relative gradients between the different fronts may also be accentuated,
with important physical consequences. Involved though it is, such an 
investigation is on our immediate agenda.

\acknowledgements{ R.M. thanks CSIR for financial support. A.B. thanks CSIR 
for support through the project 03(1074)/06/EMR-II. S.K.G. and S.R, in 
particular, thank DST for support under the IRHPA scheme.}


\begin{thebibliography}{99} 
\bibitem{key-1} E. Witten, {\it Phys. Rev.} {\bf{D30}}, 272 (1984)
\bibitem{key-2} E. Farhi and R. L. Jaffe, {\it Phys. Rev.} {\bf{D30}}, 2379 
(1984)
\bibitem{key-3} A. Bhattacharyya, S. K. Ghosh, M. Hanauske and S. Raha, 
{\it Phys. Rev.} {\bf{C71}}, 048801 (2005)
\bibitem{key-4} A. Bhattacharyya,  S. K. Ghosh and S. Raha, {\it Phys. Lett.} 
{\bf{B635}}, 195 (2006)
\bibitem{key-4a} A. Bhattacharyya and S. K. Ghosh, {\it Mod. Phys. Lett.} 
{\bf{A22}}, 1019 (2007)
\bibitem{key-5} C. Alcock, E. Farhi and A. Olinto, {\it Astrophys. J.} 
{\bf{310}}, 261 (1986)
\bibitem{key-6} N. K. Glendenning, {\it Nucl. Phys. (Proc. Suppl.)} {\bf{B24}}, 
110 (1991);  {\it Phys. Rev.} {\bf{D46}}, 1274 (1992)
\bibitem{key-7} A. Olinto, {\it Phys. Lett.} {\bf{B192}}, 71 (1987); 
{\it Nucl. Phys. (Proc. Suppl.)} {\bf{B24}}, 103 (1991)
\bibitem{key-8} M. L.Olesen and J. Madsen, {\it Nucl. Phys. (Proc. Suppl.)} 
{\bf{B24}}, 170 (1991)
\bibitem{key-9} H. Heiselberg, G. Baym  and C. J. Pethick, {\it Nucl. Phys. 
(Proc. Suppl.)} {\bf{B24}}, 144 (1991)
\bibitem{key-10} G. Lugones, O. G. Benvenuto and H. Vucetich,  {\it Phys. Rev.} 
{\bf{D50}}, 6100 (1994)
\bibitem{key-11} O. G. Benvenuto, and J. E. Horvarth, {\it Phys. Lett.} 
{\bf{B213}}, 516 (1988)
\bibitem{key-12} O. G. Benvenuto, J. E. Horvarth and H. Vucetich, 
{\it Int. J. Mod. Phys.} {\bf{A4}}, 257 (1989); O. G. Benvenuto and J. E. 
Horvarth, {\it Phys. Rev. Lett.} {\bf{63}}, 716 (1989) 
\bibitem{key-13} H. T. Cho, K. W. Ng and A. W. Speliotopoulos, {\it Phys. Lett.} 
{\bf{B326}}, 111 (1994)
\bibitem{key-14} I. Tokareva, A. Nusser, V. Gurovich and V. Folomeev, {\it Int. J. 
Mod. Phys.} {\bf{D14}}, 33 (2005) 
\bibitem{key-15} Z. Berezhiani, I. Bombaci, A. Drago, F. Frontera and A. Lavagno, 
{\it Astrophys. J.}  {\bf{586}}, 1250 (2003) 
\bibitem{key-16} I. Bombaci, I. Parenti and I. Vidana, {\it Astrophys. J.} 
{\bf{614}}, 314 (2004)
\bibitem{key-17} A. Drago, A. Lavagno and G. Pagliara, {\it Phys. Rev.} 
{\bf{D69}}, 057505 (2004)
\bibitem{key-18} A. Bhattacharyya, S. K. Ghosh, P. Joarder, R. Mallick and 
S. Raha, {\it Phys. Rev.} {\bf{C74}}, 065804 (2006)
\bibitem{key-19} J. Alam, S. Raha and B. Sinha, {\it Phys. Rep.} 
{\bf{273}}, 243 (1996)
\bibitem{key-20} G.B. Cook, S.L. Shapiro and S.A. Teukolsky, {\it Astrophys. J.} 
{\bf{422}}, 227 (1994)
\bibitem{key-21} H. Komatsu, Y. Eriguchi and I. Hachisu, {\it Mon. Not. R. Astron. Soc.} 
{\bf{237}}, 355 (1989)
\bibitem{key-23} N. Stergioulas and J. L. Friedman, {\it Astrophys. J.} 
{\bf{444}}, 306 (1995)
\bibitem{key-24} A. M. Gleeson and S. Raha, {\it Phys. Rev.} {\bf{C26}}, 
1521 (1982)
\bibitem{key-25} A. M. Anile, \textsl{Relativistic fluids and Magneto-fluids : 
with application in Astrophysics and Plasma Physics}, Cambridge University 
Press, U.K. (1989)
\bibitem{key-26} M. Gyulassy, K. Kajantie, H. Kurki-Suonio and L. McLerran, 
{\it Nucl. Phys.} {\bf{B237}}, 477 (1984)
\bibitem{key-27} Y. Raizer and Y. Zeldovich,  \textsl{Physics of shock waves and 
high temperature hydrodynamic phenomena}, Academic Press, New York (1967)
\end{thebibliography}
\end{document}